# On Capturing Laminar/Turbulent Regions Over a Wing Using WMLES

P. Balakumar
*Flow Physics and Control Branch, NASA LaRC, MS 170, Hampton, VA 23681-2199*
Email: *ponnampalam.balakumar-1@nasa.gov*

Prahladh S. Iyer
*Analytical Mechanics Associates, Hampton, VA 23666*

**Wall-modeled large-eddy simulation (WMLES) is performed for flow over a wing with a focus on documenting grid resolution requirements to predict both the laminar and turbulent regions accurately. Flow over a spanwise extruded NACA0012 airfoil at 0-degree angle of attack and freestream chord-based Reynolds numbers of 3 million is simulated using an unstructured-grid finite-volume solver. An equilibrium wall function with the first off-wall grid point as the exchange location is used. Two scenarios are simulated wherein either the entire airfoil surface, or only a portion of it, is assumed turbulent based on linear stability calculations. For the latter scenario, the regular no-slip wall boundary condition is imposed in laminar regions. Using the same grid that was employed in the fully turbulent case, WMLES captured the skin friction in the turbulent region close to the RANS results. However, the skin friction is not well captured in the laminar region, due to far few grid points inside the boundary layer. When the near-wall grid was refined in the wall normal direction, the laminar region was captured accurately, but the turbulent region was not resolved well because the first off-wall point moved below the buffer-layer. To reconcile differing resolution requirements in the laminar and turbulent regions, a grid was generated based on the varying boundary layer thickness estimated from a precursor RANS simulation with a specified transition location. This grid produced improved results in the laminar region but resulted in a delayed transition location. Unsteady disturbances with the most amplified frequencies were introduced upstream of the neutral point. The transition process in the laminar region and the skin friction in the turbulent region were then captured satisfactorily.**

## I. Introduction

Eddy-resolving methods such as direct numerical simulations (DNS), wall-resolved Large-Eddy Simulations (WRLES), and wall-modeled Large-Eddy Simulations (WMLES) are becoming increasingly adopted due to their accuracy in predicting wall-bounded turbulent flows. The simplest and the most popular approximate method is the set of Reynolds-averaged Navier-Stokes (RANS) equations, where the equations are derived for time-averaged quantities. In general, existing RANS models provide good results compared to experiments for attached flows. DNS and WRLES methods are computationally too expensive for flows over complex geometries and at high Reynolds numbers, due to the large grid resolution required near the wall. WMLES remedies this problem by modeling the inner layer of the boundary layer and resolving only the outer part of the boundary layer. Piomelli et al.[1], Spalart[2], Frohlich et al.[3], and Larsson et al.[4] review the existing WMLES and hybrid LES/RANS approaches.

There are basically two strategies currently used for WMLES. One is the hybrid LES/RANS method where the inner layer is modeled by a RANS model and the outer part is resolved by the LES.[2,3] The second is to model the inner region separately using a RANS model, to extract the wall shear stress, and to impose this wall shear stress as a boundary condition for the LES[5-8], where the LES domain extends to the wall. In this work, the latter approach is used to compute the flow over an NACA-0012 airfoil.

Flow over wings exhibits different flow dynamics and aerodynamic characteristics depending on the shape of the wings and the freestream parameters.[1,2] The freestream parameters that influence the flow dynamics are the Mach number, unit Reynolds number, and angle of attack. The shape and the angle of attack basically determine the pressure distribution on the airfoil. The pressure distribution on the pressure side (lower surface) of the wings is favorable, and the pressure distribution on the suction side (upper surface) is favorable near the leading-edge region and becomes adverse further downstream. The adverse pressure gradient tends to induce early boundary-layer transition and/or flow separation near the trailing edge of the airfoil. The transition location moves upstream closer to the leading edge with increasing angles of attack and Reynolds numbers. Tripping devices are generally used to fix the boundary-layer transition near the leading edge of an airfoil in wind tunnel experiments. This makes the flow over the airfoil fully turbulent starting from the leading edge and enables the experimental data to be extrapolated to flight conditions. In natural transition, the boundary layer remains very thin in the laminar region and increases by an order of magnitude when the flow transitions to turbulence. The question is how does WMLES perform in a fully turbulent case and in a laminar-transition-turbulent case? In the WMLES work of Refs. 6 and 9, a sensor of the amplitude of the fluctuations was used to activate the WMLES downstream of the transition location. They obtained reasonable results for the flow over a multi-element airfoil. In another approach,[10, 11] nonlinear parabolized stability equations were first used to compute the evolution of linear and weakly nonlinear disturbances inside the boundary layer up to the skin friction rise. These disturbances and the mean flow are then fed into the inflow of the WMLES domain. Good results were obtained in simulations over a flat plate. Recently, WMLES were performed for flow over an airfoil with laminar and turbulent regions.[12] Different approaches were attempted to resolve the laminar region: (i) apply the wall model everywhere, (ii) change the exchange location to the first cell in the laminar region, (iii) assume a linear velocity profile near the wall, and (iv) obtain the wall stress in the laminar region from a boundary-layer solution. The results showed significant differences in lift and drag among the different approaches.

WMLES with equilibrium and nonequilibrium wall-models has been successfully applied to plane incompressible and compressible channel flows,[5, 7, 9] flat-plate boundary layers,[6, 8, 9] flow over an NACA-0014 airfoil,[13] the NASA hump,[14, 15] and a multielement airfoil[5]. Recently, the WMLES methodology has been successfully used to simulate flows over semi- and full-span aircraft configurations near flight Reynolds numbers, as part of the Fifth AIAA High-Lift Prediction Workshop.[16, 17] The computed lift, drag, and pitching moments agree well with experimental data. However, the state of the boundary layer near the leading-edge region of the airfoil, whether laminar, turbulent, or transitional, introduces uncertainties in predictions of aerodynamics quantities. Another observation in these simulations is that the lift, drag, and moments are dominated by pressure forces; the viscous forces contribute less than 5% of the total drag force at high-lift conditions.

As a prelude to predict turbulent flows accurately over 3D geometries, the objective of this work is to evaluate the predictive capabilities of WMLES for the flow over a NACA-0012 airfoil at 0 degree angle of attack, which is dominated by viscous forces and the results have a strong sensitivity to the transition location and wall model. A secondary objective is to document the grid requirements for the laminar and turbulent regions of the flow and suggest improved grid-generation, and modeling practices for simulating such flows accurately in the WMLES framework. The problem setup and numerical details are described in Sections II, precursor RANS and linear stability results are discussed in Sections III and IV, and WMLES results are described in Section V. Finally, summary and conclusions from this study are presented in Section VI.

## II. Numerical Details and Problem Description

WMLES is performed using the FUN3D code, which is a node-based, finite-volume code developed at NASA's Langley Research Center that solves the three-dimensional compressible Navier-Stokes equations on unstructured grids.[18, 19] The implementation of WMLES methodology in FUN3D is described in Wang et al. (2023).[20] The Spalding log-law is used as the wall model, with the first grid point off the wall as the exchange location in the wall-stress calculations. Periodic conditions are applied in the spanwise direction, and characteristic boundary conditions are used at the outer boundary. A spanwise extent of 0.05c and an outer boundary of radius $R_c = 100c$ are used. The grid is uniform in the spanwise direction.

Computations are performed for a flow over an NACA-0012 airfoil with a sharp trailing edge.[21] The chord length, *c*, is used to nondimensionalize the coordinates. The airfoil has a zero thickness at the trailing edge. A Cartesian coordinate system with the *x*-axis along the chord, *y*-axis along the span, and *z*-axis perpendicular to the chord is used. The coordinates are nondimensionalized by the chord length, *c*. The corresponding velocity components are denoted by *u, v,* and *w*, respectively. The variables density $\rho$, pressure *p*, and the velocities are nondimensionalized by their corresponding freestream variables $\rho_\infty$, $p_\infty$, and $U_\infty$, respectively. The time is nondimensionalized by $(c/U_\infty)$. Simulations are performed at a Reynolds number of $Re_c = \rho_\infty U_\infty c / \mu_\infty = 3 \times 10^6$, and at an angle of attack of 0 degree. The freestream Mach number is $M = 0.3$. Experiments for the NACA-0012 airfoil were performed[22] in the Langley Research Center 8-foot Transonic Pressure Tunnel for a range of Mach numbers, from 0.30 to 0.86, and for Reynolds numbers of 3.0 million and 9.0 million.

WMLES is performed for a fully turbulent case wherein the entire airfoil surface is assumed to be turbulent, and a wall model is applied. The WMLES results are compared with WRLES and RANS results. To investigate the laminar-turbulent flow scenario, linear stability computations are first performed to determine the frequencies and the N-Factors of the most amplified instability waves. The transition onset locations are then extracted based on a critical N-Factor of 9. RANS simulations are performed for later comparisons with the WMLES results, assuming laminar flow upstream of the transition front and turbulent downstream. Simulations are conducted by imposing no-slip conditions upstream of the transition front and wall-model conditions (WMLES) downstream. Computations are performed using two grid systems. In the full turbulent case, a structured C-type was used. The grid system used is shown in Fig. 1(a). The first cell off the wall has a constant height across the wing. The first grid point is located in the log-layer except in the leading-edge region. Here, the grid is very coarse and not able to resolve the thin laminar boundary layer upstream of the transition point. Additional simulations are also performed by reducing the grid sizes off the wall. To remedy this for the laminar-turbulent WMLES, an unstructured grid was generated with fixed number of points inside the boundary layer over the wing. This is done using the boundary-layer thickness variations extracted from the RANS simulations. Since the boundary-layer thickness increases by several folds across the laminar-turbulent transition region, the grid distribution normal to the surface also changes accordingly. The unstructured grid distributions are depicted in Figs. 1(b, c). The grids for WMLES have an outer boundary of 100c, where characteristic boundary conditions are imposed, and are extruded along the span of 0.05c with a uniform spacing of 0.0005c.

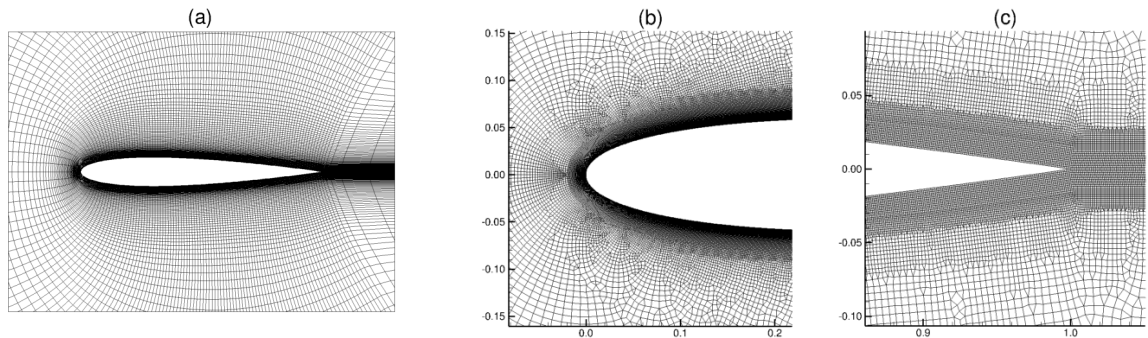

**Figure 1. Grid distributions used in WMLES. (a) Structured (every 20$^{th}$ point is shown in the streamwise direction) and (b, c) unstructured.**

## III. Precursor RANS Simulations

We first perform RANS simulations using FUN3D and the SA turbulence model on a wall-resolved grid with $\Delta z^+_w < 1$. Simulations are performed assuming fully turbulent flow and specifying laminar flow upstream of the transition front and turbulent downstream. The transition front is selected to be at $x_{tr} = 0.40$ and is slightly downstream of the transition onset point of 0.36 predicted from the N-Factor results. The laminar boundary layer at this angle of attack separates around $x \sim 0.60$. Figures 2 (a) and (b) depict the pressure and the skin friction distributions, respectively, for the fully turbulent case and the laminar-turbulent case with $x_{tr} = 0.40$. The skin-friction shows the laminar variations upstream of the transition front ($x = 0.40$) and turbulent variations downstream. The pressure distributions are almost the same for all the cases except small dip near the transition onset point. Figures 3(a) and (b) show boundary layer profiles at various stations $x = 0.2, 0.4, 0.6, 0.8, 0.9$ for the the fully turbulent and laminar-turbulent cases, respectively. Figure 3(a) shows the boundary layer profiles for the fully turbulent case. Here, $z_n$ is the normal distance from the surface. The corresponding boundary-layer thicknesses are (0.004, 0.008, 0.013, 0.017, 0.022). We marked 100, 10, and 1 percent of the boundary layer thickness in these plots to elucidate the extent of the log region inside the boundary layers. The ten percent boundary-layer height varies from $4 \times 10^{-4}$ to $2 \times 10^{-3}$. Figure 3(b) depicts the boundary layer profiles in the laminar region at stations $x = 0.2, 0.4$ and in the turbulent region at $x = 0.6, 0.8, 0.9$. The corresponding boundary-layer thicknesses are (0.0018, 0.0023, 0.0072, 0.0124, 0.0144). Figure 4 displays the variations of the boundary layer thickness in semi-log scale for the fully turbulent and the laminar-turbulent cases. We also marked the laminar (L) and the turbulent (T) regions in the figure. The boundary-layer thickness increases by about 2.3 times from $x = 0.30$ to $x = 0.50$.

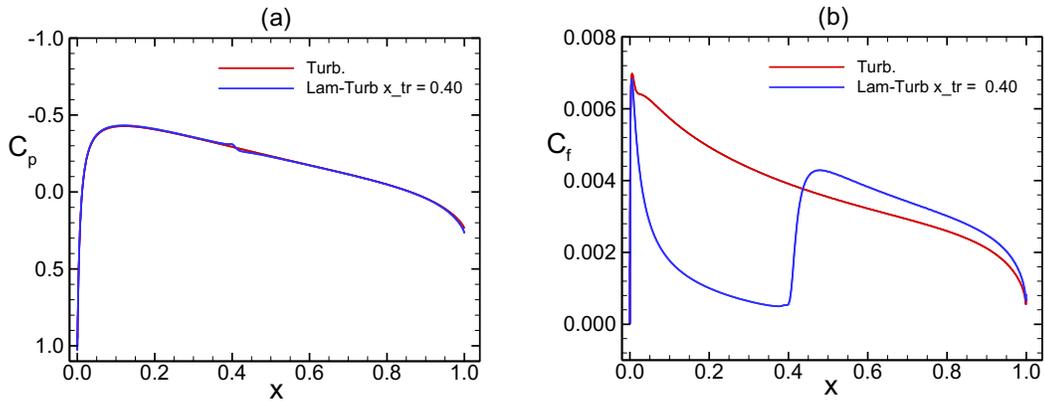

**Figure 2. Mean quantities obtained from RANS computations with the transition onsets at $x_{tr} = 0.40$. (a) Pressure coefficients and (b) skin friction coefficients.**

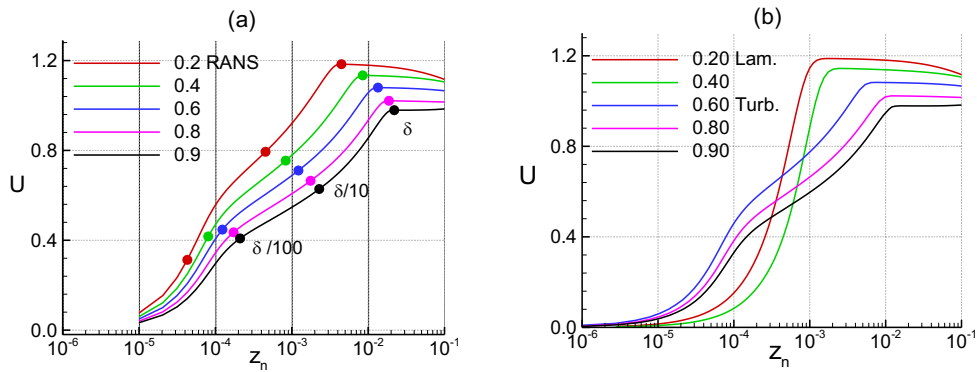

**Figure 3. Boundary layer profiles obtained from RANS computations (a) fully turbulent and (b) with transition onset at $x_{tr} = 0.40$ at $x = 0.2, 0.4, 0.6, 0.8, 0.9$.**

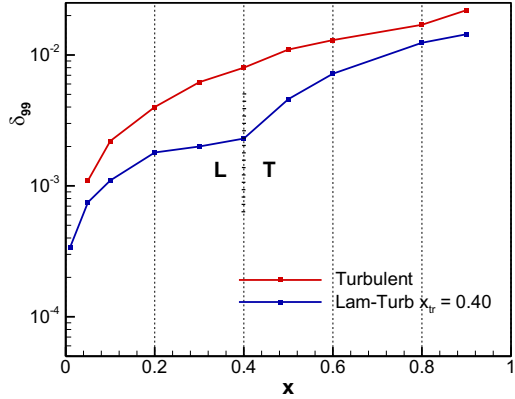

**Figure 4. Variations of the boundary layer thickness for the fully turbulent and the laminar-turbulent case with $x_{tr} = 0.40$.**

## IV. Linear Stability Calculations

The coordinates of the transition front are needed to separate the laminar and turbulent zones for the WMLES computations. This can be obtained from experiments or from any transition prediction tools, such as linear stability and the N-Factor method.[23] In this work, linear stability and N-Factor methods are used to locate the transition onset points on the wing. A panel method is used to get the inviscid pressure distributions on the surface and a boundary layer code is used to obtain the boundary-layer velocity profiles. Linear stability and N-Factor calculations are performed using an in-house stability code.

Figure 5(a) displays the N-factor curves for the $Re_c = 3 \times 10^6$. The curves are shown for different nondimensional frequencies $F_0 = 2\pi \nu_\infty f / U_\infty^2$. Here, $f$ is the dimensional frequency in $Hz$, $U_\infty$ and $\nu_\infty$ are freestream velocity and kinematic viscosity in the freestream, respectively. An N-Factor of 9 is used to fix the transition onset in the computations. The predicted transition onset point is 0.36. The results at this Reynolds number agree with the results of Halila et al.[24] The most amplified frequency is $F_0 = 6.5E-5$. The wavelengths of the most amplified frequencies are about 0.011. To validate the existence of these instability waves near the leading-edge region of the wing, two-dimensional direct numerical simulations are performed using a compact scheme.[25, 26] Disturbances with the most amplified frequencies are introduced upstream of the corresponding neutral points by applying localized harmonic normal velocities of small amplitudes at the wall. Simulations are also conducted without any external forcing to study the types of disturbances generated inside the boundary layer over the wing in natural numerical environments. Figure 5(b) shows instantaneous pressure variations along the wing surface. The results with and without forcing are shown in blue and red colors, respectively. First observation is that with the forcing at the most amplified frequencies, disturbances grow exponentially starting close to the neutral points as predicted by the linear stability computations. It is also noted that without any forcing, disturbances start to appear further downstream than predicted by the linear theory. It is seen that the disturbances start to grow from $x \sim 0.10$ with the forcing and appear near $x \sim 0.60$ without forcing. One implication is that if a sensor-based method is used to detect the starting location of WMLES, the sensor might detect a station much further downstream than the actual transition location.

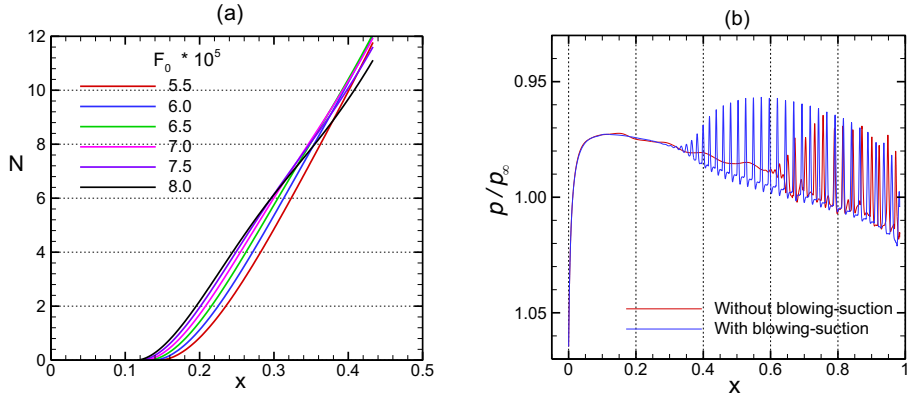

**Figure 5. (a) N-Factor curves obtained from linear stability calculations and (b) instantaneous pressure distributions along the surface.**

## V. WMLES Results

*(a) Fully turbulent case*

We present the results obtained from the WMLES assuming the flow is fully turbulent from the leading edge of the wing. A grid size of (4001, 141, 101) was used along the airfoil surface, in the wall normal direction, and in the spanwise direction, respectively. A C-type structured grid was employed (Fig. 2(b)). This corresponds to a nominal grid spacing of $3 \times 10^{-4}$ and $5 \times 10^{-4}$ in the tangential and spanwise directions, respectively. The wall-normal spacing closest to the airfoil surface is located approximately in the middle of the log-layer at $x = 0.2$ and is in the lower part of the log-layer at $x = 0.9$. Figure 6(a) displays the instantaneous $u$-velocity contours in ($x$-$z$) planes. The grid distributions are also included in the figures. It is seen that the boundary layer is very thin near the leading edge, where there are only one or two points inside the boundary layer. The number of points inside the boundary layer increases to about 6 and 20 points near $x = 0.15$ and 0.8, respectively. WRLES calculations were also performed at this Reynolds number of $Re_c = 3 \times 10^6$ and angle of attack of $\alpha = 0°$ using a grid size of (12001, 257, 501). For these calculations, unsteady body forces are applied near $x \sim 0.05$ to trip the boundary layer.

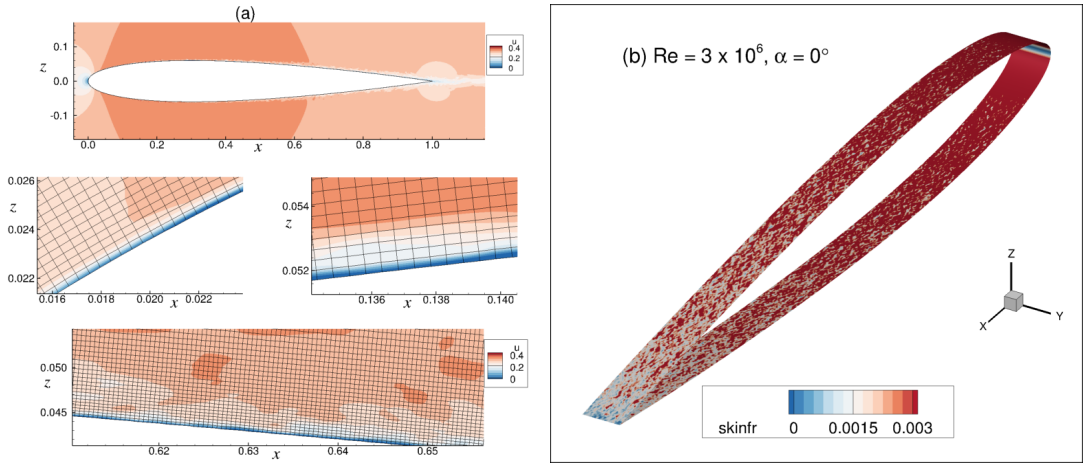

**Figure 6. Instantaneous (a) $u$-velocity contours and grid distribution in the ($x$-$z$) plane at $y = 0$ and (b) skin friction distribution for WMLES.**

Figure 6(b) depicts contours of instantaneous skin friction coefficient on the surface of the wing. We applied the wall model for the entire airfoil starting from the leading edge. The figure shows that the unsteady turbulence structures begin to form very close to the leading edge. The N-factor computations predicted the natural laminar-to-turbulent transition will occur around $x = 0.4$. In the absence of tripping or external disturbances, WRLES will transition further downstream of the expected natural transition locations. However, it is interesting that boundary layer breaks down to turbulence very close to the leading-edge, despite the absence of external tripping. The physical mechanism behind this tripping needs to be understood further, to potentially shed light on behavior in attached regions in more complex configurations.

Figures 7(a, b) show the time-averaged wall pressure and skin friction coefficients obtained from the WMLES, WRLES, and RANS for $Re_c = 3 \times 10^6$. The pressure coefficient for WMLES and WRLES is almost the same as the RANS results. The skin-friction coefficient is lower near the leading-edge region compared to the RANS results, and further understanding of this behavior is necessary. This is presumably related to the laminar-turbulent tripping mechanism. As discussed previously, the grid is very coarse in this region and the boundary layer is not resolved well. The skin-friction is lower than the RANS results, but the trend is the same. Figure 9 displays the velocity profiles in log scale at stations $x = 0.2$, 0.6, and 0.8. The vertical black line in these figures indicates the WMLES exchange locations. The exchange locations are above the buffer layer and in the lower part of the log-layer. The magnitudes of the velocities are lower than RANS in the buffer region and higher in the outer part of the boundary layer. The agreement with the RANS results is reasonable.

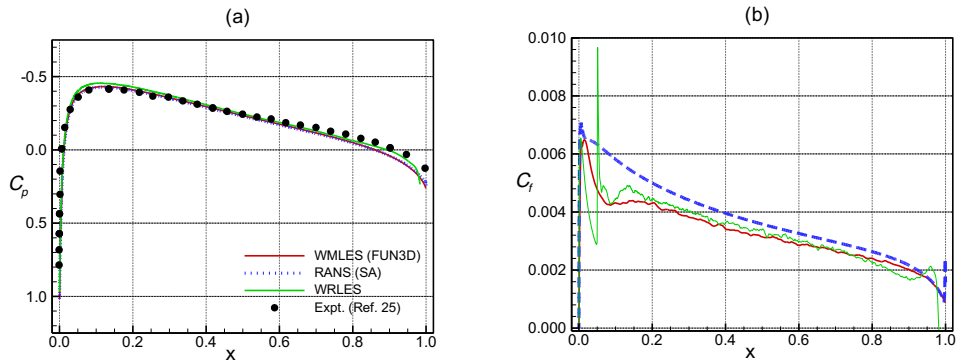

**Figure 7. Mean (a) pressure and (b) skin friction coefficients for the fully turbulent case.**

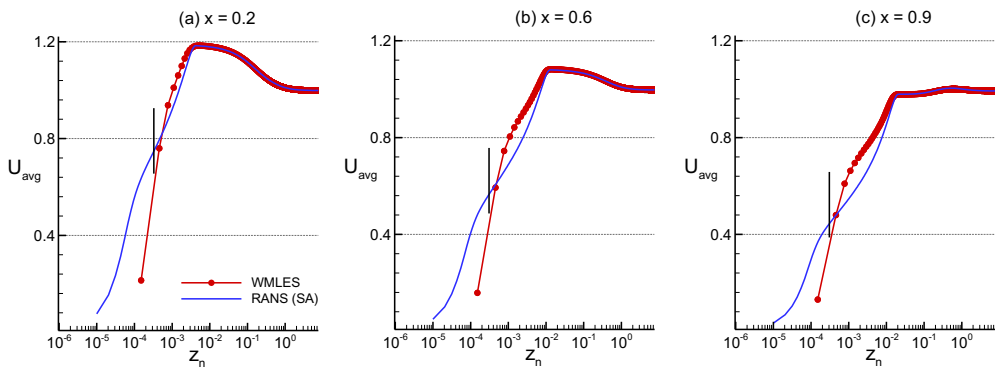

**Figure 8. Mean velocity profiles in log scale at $x = 0.2$, 0.6, and 0.9 for the fully turbulent case. Vertical lines denote the exchange locations for WMLES.**

Figures 9(a-c) depict the computed mean boundary layer profiles in linear scale at $x = 0.2$, 0.6, and 0.9 from the WRLES and WMLES for $Re_c = 3 \times 10^6$. The RANS results are also included for comparison. The mean velocity profiles obtained from RANS agree very well with the WRLES results. This validates the SA turbulence model in predicting the turbulent profiles over airfoils at low angles of attack and confirms the effectiveness of the tripping technique employed in the simulation. The WMLES result near the leading edge at $x = 0.2$ agrees with WRLES and RANS, but further downstream the velocity profiles deviate from the WRLES and RANS results. The WMLES velocities are larger in the outer part of the boundary layer. Similarly, Figure 10 depicts the results for the streamwise normal Reynolds stresses $<u'u'>$. The intensity of the $<u'u'>$ fluctuations near the wall are under predicted by RANS. However, the agreement is good in the outer part of the boundary layer. WMLES better predicts $<u'u'>$ compared to RANS, especially at $x = 0.2$ and $x = 0.6$. It should be noted that the RANS simulation and WMLES were performed with a sharp trailing edge, while WRLES was performed with a blunt trailing edge. This could be responsible for some of the observed discrepancies, especially close to the trailing edge. For instance, the velocity profiles at $x = 0.2$ and $x = 0.6$ agree very well between RANS and WRLES and begin to deviate further downstream. Also, a C-type grid was used for the WMLES, which might explain some differences. These results imply the fully turbulent case can be captured by the WMLES accurately.

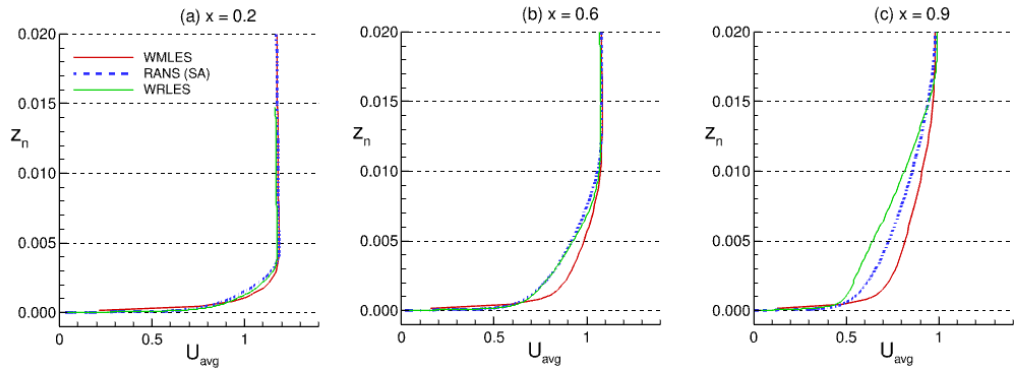

**Figure 9. Time-averaged velocity profiles at $x = 0.2$, 0.6, and 0.9 for WRLES, WMLES, and RANS.**

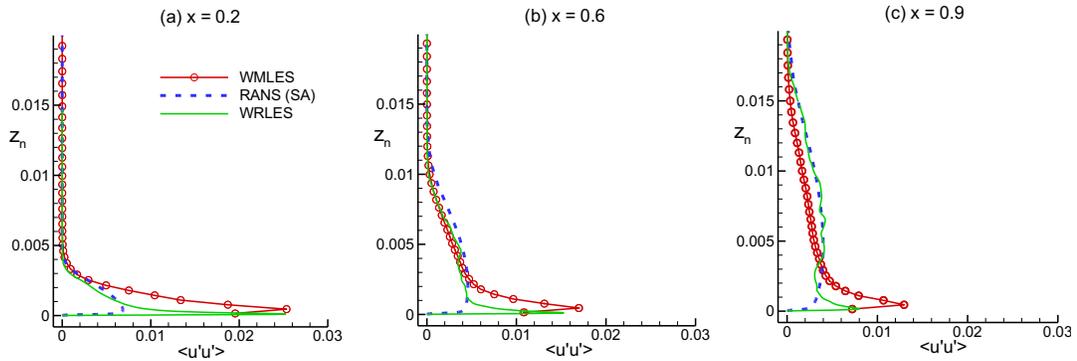

**Figure 10. Normal Reynolds stress $<u'u'>$ profiles at $x = 0.2$, 0.6, and 0.9 for WRLES, WMLES, and RANS.**

*(b) Laminar- turbulent case*

In this section, the results obtained from the WMLES for the laminar-turbulent case are presented. This is a more difficult case to predict than the fully turbulent case. The boundary layer thickness is small in the laminar region, and it increases significantly downstream the laminar-turbulent transition. Another issue is that in WMLES, it is necessary to place the first grid point off the wall in the log-layer due to lack of resolution needed to resolve the wall dynamics below the buffer layer. These features make it difficult to generate a grid topology to resolve both laminar and turbulent regions accurately. The coordinates of the transition front are also needed to separate the laminar and turbulent zones for the WMLES computations.

This can be obtained from experiments or from any transition prediction tools, such as linear stability and the N-Factor method.[23]

WMLES is first conducted employing the same grid used in the fully turbulent case (Fig. 1(a)), that is a C-type structured grid with the first grid point located approximately 3 x $10^{-4}$ above the surface. The grid point (3 x $10^{-4}$) is in the lower part of the log-region, as seen in Fig. 3(a). Figure 11 shows contours of the instantaneous skin-friction coefficient obtained from the WMLES for different $\Delta z_n$ = 3 x $10^{-4}$ and 3 x $10^{-5}$. Figures 12(a) and (b) depict the time-averaged skin-friction and pressure coefficients, respectively. In these figures, results obtained from the RANS calculations are included for comparison. It is interesting to observe for the case with 3 x $10^{-4}$ that the flow remains laminar up to the transition front $x$ = 0.4 and enters a turbulent regime immediately downstream. The time-averaged skin-friction variation shows that the shape in the laminar region follows the steady laminar-turbulent RANS results, despite the coarser grid distribution in this region. The amplitudes of the skin-friction are underpredicted until $x$ = 0.10, compared to the laminar-turbulent results and slightly overpredicted until $x$ = 0.4. The skin friction distribution in the turbulent region is very close to the RANS results. The pressure coefficient agrees well with the RANS prediction despite small over and under predictions in the laminar and turbulent regions, respectively. These are encouraging results in the sense that if small discrepancies in the laminar region can be accepted, the overall WMLES predictions are reasonably good. The only information needed for these simulations is the coordinates of the transition locations. As described earlier, this can be obtained from experimental data or from numerical transition prediction methods. To investigate what kinds of grid distributions are needed to capture the laminar region accurately, additional simulations were carried out by decreasing the heights of the first cell to 2 x $10^{-4}$, 1 x $10^{-4}$, and 3 x $10^{-5}$. Figure 11(b) displays the instantaneous skin-frictions contours obtained. There are about 10 grid points inside the laminar boundary layer near the leading-edge with the smallest grid size of 3 x $10^{-5}$. As expected, the skin-friction and pressure coefficients are very well captured in the laminar region using this grid. However, the turbulent region is not captured correctly. The boundary layer transitioned to turbulence further downstream at $x$ = 0.6 instead of $x$ = 0.4. This is due to the first grid point being located below the buffer layer and the spanwise grid distribution being kept the same, such that the wall model is not able to trigger the transition to turbulence near $x$ = 0.4. With decreasing heights of the first grid point, the laminar region is resolved better. The grid with 1x$10^{-4}$ resolves the laminar part accurately. This suggests that about 4-5 points in the leading-edge region are sufficient to resolve the laminar boundary layer accurately. However, the turbulent region is not captured well except for the 3 x $10^{-4}$ grid. This suggests that to correctly capture the turbulent region, the first grid point should be above the buffer layer and in the middle of the log-layer.

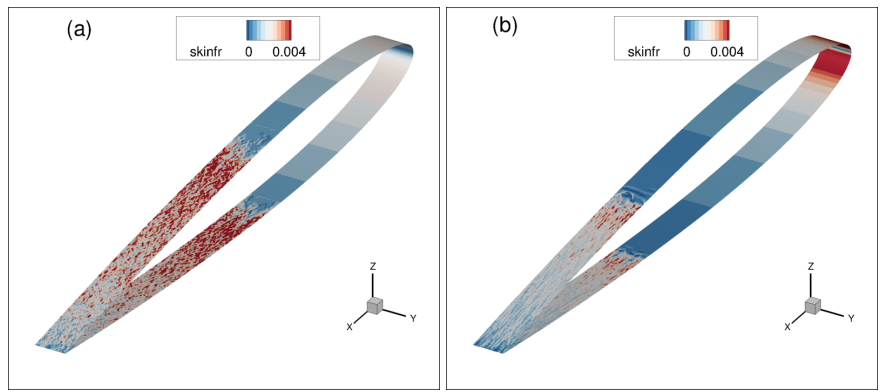

**Figure 11. Contours of instantaneous skin friction coefficient with (a)** $\Delta z_n = 3 \times 10^{-4}$ **and (b)** $\Delta z_n = 3 \times 10^{-5}$.

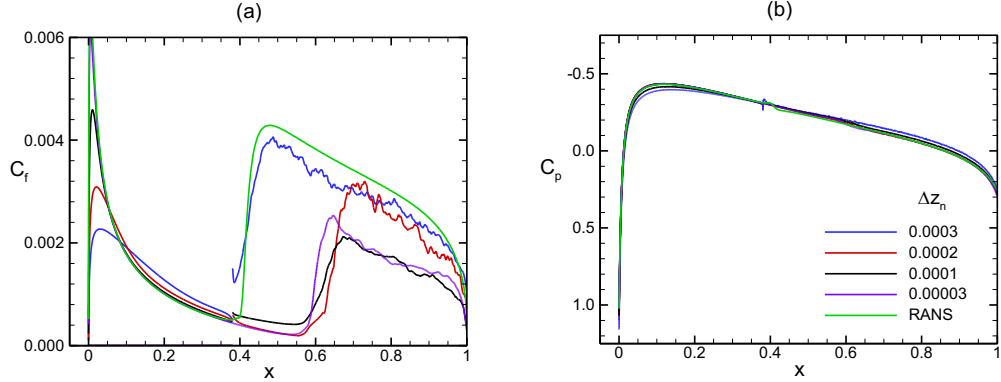

**Figure 12. Time-averaged skin-friction coefficients obtained from WMLES for different $\Delta z_n$.**

The simulation results from previous sections show the difficulties in resolving the laminar and turbulent regions with grids generated with constant first grid sizes off the wall. If the grid is coarse, WMLES will capture the turbulent region reasonably well if the transition front is known *a priori*, but the laminar region will not be captured accurately. If the grid is finer near the wall, the results show the opposite behavior with the laminar region resolved well and the turbulent region captured poorly. Another observation is that even though there exist very strong boundary-layer instability waves in the laminar region, no instability waves are observed in this region in any of the simulations. It should also be noted that the WMLES grids are very coarse by design and are not suitable for capturing instability waves accurately. This will be an issue when one uses sensor-based approaches to decide on which zones wall models should be applied. In this paper, two approaches were attempted to resolve these issues. One is to design a boundary layer grid based on the variation of the boundary-layer thickness along the wing. Using this approach, a fixed number of points can be assigned in the laminar as well as in the turbulent boundary layers. The second is to introduce external disturbances with the most amplified frequencies near the leading edge. The evolution of the disturbances may not be captured very accurately with coarse grids, but with large initial amplitudes, one may be able to generate instability waves inside the boundary layer. Another feasible approach, which was not considered in this work, is to explore overset grid topologies.

An unstructured grid is generated with about 20 points per boundary-layer thickness (ppd). The grid has about 4000 points distributed equally around the wing, and 100 points distributed uniformly across the span of width 0.05 (see Figs. 1(b, c)). A two-dimensional harmonic body force of Gaussian shape in *x* with a frequency of $F_0 = 5 \times 10^{-5}$ is applied inside the boundary near the leading edge, at $x = 0.05$. The body force is applied only on the upper side of the wing. The wavelength of the most amplified frequency is about 0.01. Hence there are about 20 points per wavelength in this grid distribution. Figures 13(a) and (b) show the instantaneous skin-friction coefficient on the surface of the wing in three-dimensional and in two-dimensional plan views, respectively. Figures 14(a) and (b) display the instantaneous and time-averaged skin-friction coefficients and Fig. 14(c) shows the time-averaged pressure coefficient. Firstly, the laminar region shows excellent agreement with the RANS results due to sufficient grid points in the wall-normal direction within the boundary layer. Next, turbulence is observed earlier on the upper surface than on the lower surface even though the grid distributions are symmetric about the *x*-axis, indicating the importance of explicit tripping for the chosen grid resolution. Growing two-dimensional waves and their breakdown to three-dimensional disturbances are seen near $x = 0.4$ in Fig. 13 and Fig. 14(a). Figure 15 displays the mean velocity profiles at $x = 0.8$ obtained from WMLES using the C-grid with the first off-wall grid point at a constant height $\Delta z_n = 3 \times 10^{-4}$ (corresponds to Fig. 1(a)) and the unstructured grid with fixed number of points in the boundary layer (Fig. 1(b-c)). RANS results are included for comparison. The profiles obtained using different methods agree reasonably well with each other and with the RANS. The amplitudes of the forcing could have been adjusted to force the breakdown earlier to match the experimentally observed or numerically predicted transition front. The question of how to use a sensor-based method to select the zones to apply the wall model remains open. In this work, two concepts are tested to determine their performance, and these questions will be explored in future work.

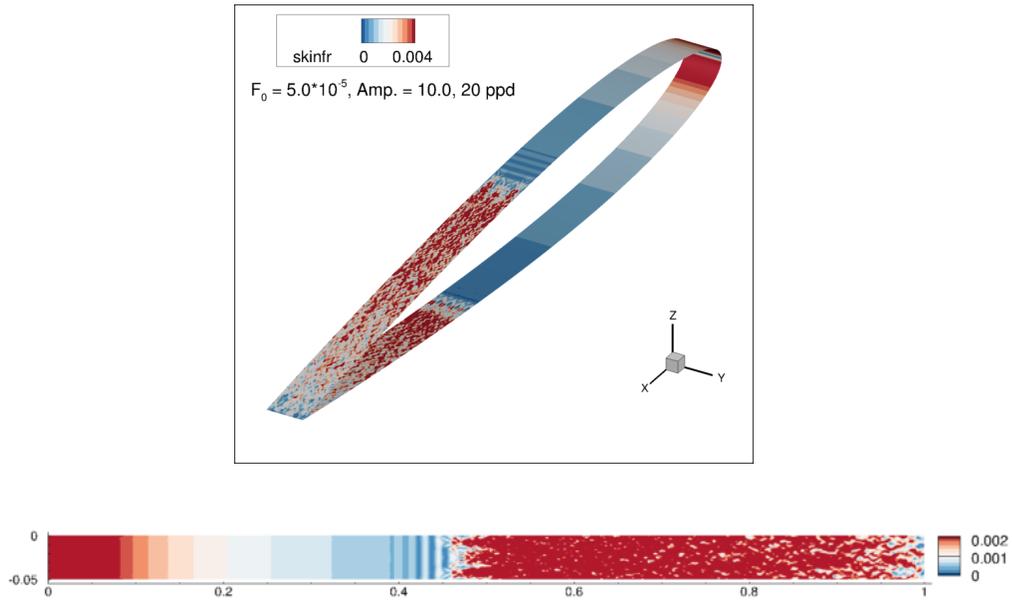

Figure 13. Contours of the instantaneous skin friction coefficient for 20 ppd grid.

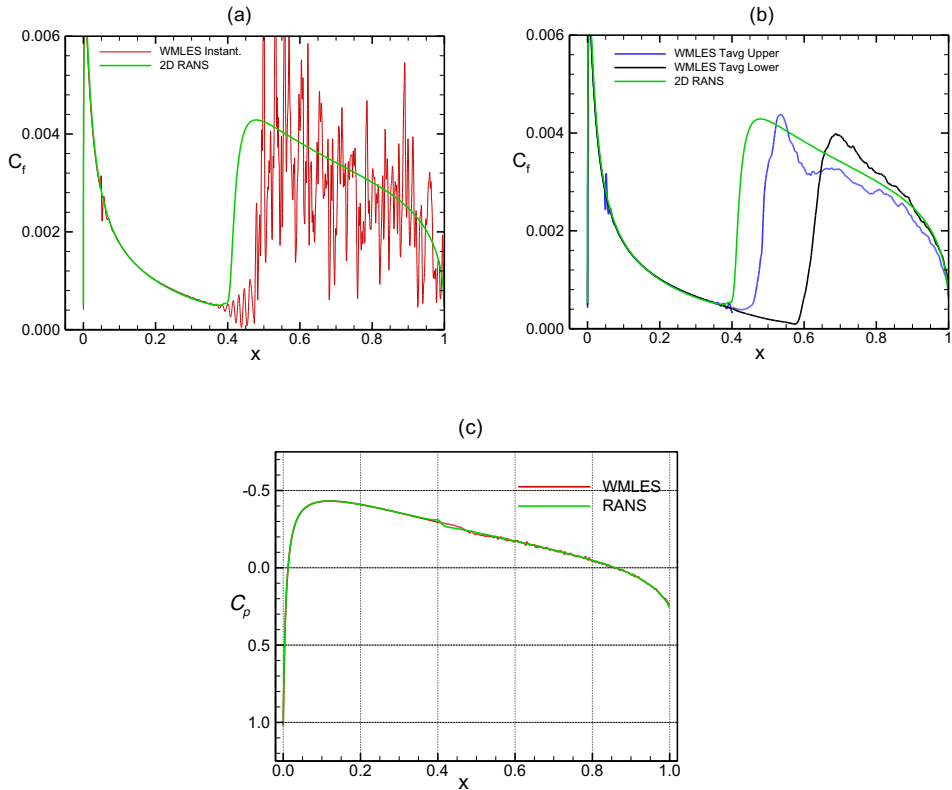

Figure 14. Simulation results from WMLES for 20 ppd grid. (a) Instantaneous skin friction coefficient, (b) time-averaged skin-friction coefficient, and (c) time-averaged pressure coefficient.

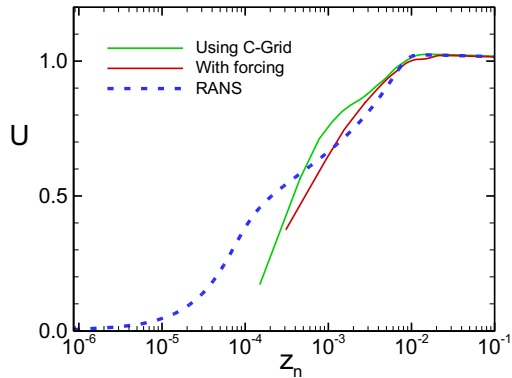

**Figure 15. Time-averaged velocity profiles in log scale at *x* = 0.8 obtained from WMLES for the laminar-turbulent case**.

## VI. Summary and Conclusions

WMLES was performed for flow over a NACA-0012 airfoil at chord Reynolds numbers of $Re_c = 3 \times 10^6$, freestream Mach number of 0.3, and 0-degree angle of attack using an equilibrium wall-model with first off-wall grid point as the exchange location. Two different flow scenarios were investigated. In one case, the flow was assumed to be fully turbulent and in the other the flow was assumed to be laminar up to a point on the wing and turbulent downstream. In the fully turbulent case, WMLES results showed turbulent flow being triggered from the leading edge and generally showed good agreement with WRLES and RANS. Generating a suitable grid for the laminar-turbulent case is challenging since it requires *a priori* knowledge about the boundary layer thickness, which varies significantly between the laminar and turbulent regions. Grids with uniform wall-normal distribution over the surface either predicted the laminar or turbulent regions well, depending on the grid spacing, but not both. Using exchange locations away from the wall in the turbulent region might produce improved predictions. On the other hand, grids with varying wall-normal distribution based on the local boundary layer thickness predict both the laminar and turbulent regions well provided a suitable tripping is present. In the absence of explicit tripping, the onset of turbulence was delayed for the grid resolution used in this study. Overall, generating grids with a wall-normal distribution based on the local boundary layer thickness and including an explicit tripping mechanism appears to be a promising strategy for capturing flows with laminar and turbulent regions, and will be explored for more complex configurations in the future.

## Acknowledgments

This research is sponsored by the NASA Transformational Tools and Technologies (TTT) project of the Transformative Aeronautics Concepts Program under the Aeronautics Research Mission Directorate. The authors would like to thank Dr. Li Wang, Dr. Gary Coleman and Dr. Stephen Woodruff from the Computational AeroSciences Branch at NASA Langley Research Center for their careful reading of the paper and valuable comments.